\documentclass[aps,floatfix]{revtex4}

\usepackage{amssymb} \usepackage{graphicx} \usepackage{amsmath}
\usepackage[T1]{fontenc} \usepackage{pstricks} \usepackage{subfigure}
\usepackage[colorlinks=true,citecolor=blue,urlcolor=red]{hyperref}
\usepackage[normalem]{ulem}
\usepackage{hyperref}
\usepackage{doi}

\newcommand{\M}{M_{\rm BH}}
\newcommand{\m}{m_{\rm wd}}

\begin{document}
\title{Modelling quantum aspects of disruption of a white dwarf star by a black hole}

\author{Tomasz Karpiuk$^{1}$}
\author{Marek Niko{\l}ajuk$^{1}$}
\author{Mariusz Gajda$^{2}$}
\author{Miros{\l}aw Brewczyk$^{1}$}

\affiliation{$^{1}$Faculty of Physics, University of Bia{\l}ystok, Cio{\l}kowskiego 1L, 
           15-245 Bia{\l}ystok, Poland\\
$^{2}$Institute of Physics, Polish Academy of Sciences, 
           Aleja Lotnik{\'o}w 32/46, 02-668 Warsaw, Poland  }

\date{\today}
\maketitle

\textbf{
  We study the final stages of the evolution of a binary system
   consisted of a black hole and a white dwarf star. We implement
   the quantum hydrodynamic equations 
  and carry out numerical simulations. As a model of a white dwarf
   star we consider a zero temperature droplet of attractively
   interacting degenerate atomic bosons and spin-polarized atomic
   fermions. Such mixtures are investigated experimentally nowadays.
   We find that the white dwarf star is stripped off its mass while
    passing the periastron.  Due to nonlinear effects, the accretion disk originated
    from the white dwarf becomes fragmented and the onset of a quantum
    turbulence with giant quantized vortices present in the bosonic component of the 
    accretion disk is observed. The binary system ends its life in a spectacular way,
    revealing quantum features underlying the white dwarf star's
    structure. We find a charged mass, falling onto a black hole,
    could be responsible for recently discovered ultraluminous X-ray
    bursts. The simulations show that final passage of a white dwarf near a black hole 
    can cause a gamma-ray burst. \\ \\
}

\noindent
White dwarf (WD) stars are ubiquitous in the Universe.  They can be found as
companions in various binary systems, including those with ordinary
stars, giant stars, or compact objects as another white dwarfs,
neutron stars (NSs), or black holes (BHs) 
\citep[e.g.][]{Napiwotzki04,Monelli05,Brown17,Kaplan14,Kepler16}.

The dynamics of a white dwarf in the field of a black hole, in
particular its tidal disruption (TD), has been modeled for years. Typically
a smoothed particle hydrodynamics (SPH) simulations are performed to
monitor the behavior of a white dwarf. In this simulations a white
dwarf is represented by a collection of SPH particles according to 
the Helmholtz equation of state.  The self-gravity of the white dwarf is 
included but with adaptive gravitational softening \citep{Price07,Dehnen12}.
The other group of numerical approaches to binary systems is based on
general relativistic hydrodynamic simulations. The hydrodynamic
equations, enriched by the polytropic equation of state of stellar
matter, has been already used to study the TD of the
main-sequence star by the BH \citep{Ryu19} or the merger of
the WD and NS \citep{Paschalidis11}.

Here, we are studying the dynamics of a model WD in the
field of a BH incorporating quantum hydrodynamics.
As a model of cold WD star we propose
to consider the Bose-Fermi droplet consisting of ultracold bosonic and
fermionic atoms. Such systems have been recently predicted
theoretically \citep{Rakshit18,Rakshit18a}.
An atomic Bose-Fermi droplet can exists
because of subtle interplay of two effects. The first one is related
to the attraction between bosons and fermions. If it is strong enough
then bosons start to effectively attract each other \citep{Chin19} and
the droplet becomes unstable against collapse. Then all particles in
the droplet attract each other just like particles in the WD
star attract themselves gravitationally. This collapse can be stopped
by the fermionic component of the system. Its quantum pressure, like
the pressure of degenerate electrons in the WD, is able
to counteract the collapse and, to some extent, stabilize the
system. 




While forming, temperatures of WDs are of the order of $10^5\,$K. 
They cool down to $\sim 3\times 10^4\,$K within ten million years.
About a billion years is needed to decrease the temperature to $10^4\,$K \citep{Fontaine01}.  
The oldest observed white dwarfs have temperatures of the order of $10^3\,$K. 
For example, the effective temperature limit of the companion of the PSR J2222-0137 is estimated
to be $\lesssim$ 3500 K \citep{Kaplan14}.  
At the same time the WDs are extremely dense systems, with the densities between
$10^4$-$10^7$g cm$^{-3}$. It results in the very high Fermi temperature, 
a few orders of magnitude larger than the WD temperatures. 
Indeed, it is save to treat electrons as a
gas at zero temperature. On the other hand, for densities
of $10^6$g cm$^{-3}$, the critical temperature for the Bose-Einstein
condensation for $^{4}$He component (neglecting interactions)
becomes $10^5\,$K making an assumption that most of bosonic component
is condensed very plausible. 

Hence, we propose here to consider
bosonic component of a WD as a gas at zero temperature as
well, although the description of bosons including thermal fraction is
already well known \citep{review}. Let us mention that the possibility
of formation of the Bose-Einstein condensation in helium white dwarf
stars was already discussed in
\cite{Gabadadze08a,Gabadadze08b,Gabadadze09,Mosquera10}. \\ \\

\newpage
\noindent
\textbf{Quantum hydrodynamic approach} \\
To describe Bose-Fermi mixtures we use the formalism of quantum hydrodynamics
\citep{Madelung}. One of the first attempts to
discuss fermionic gases within this framework was already done many
years ago, see Ref. \cite{Wheeler}, where the oscillations of
electrons in a many-electron atom induced by ultraviolet and soft
X-ray photons were studied. Here, we follow this reasoning and apply quantum
hydrodynamic equations both for fermionic and bosonic clouds in a
droplet.

The quantum hydrodynamic description is briefly presented in the Methods.
Although initially the atomic number densities for both components 
and the corresponding velocity fields are basic variables of our model,
we eventually, by using the
inverse Madelung transformation, put description of the system in terms of
the wave function for bosons, $\psi_B({\bf r},t)$, and the fermionic pseudo-wave 
function, $\psi_F({\bf r},t)$.

Now we place the Bose-Fermi droplet in the field of an artificial
black hole. We assume a non-rotating black hole described by the
Schwarzschild space-time metric. A motion of a test particle in the
Schwarzschild metric conserves both the energy and the orbital angular
momentum. The energy of a test particle can be, as in the Newtonian
case, divided into kinetic and potential energies. The latter contains
the additional, with respect to the Newtonian case, term which is
proportional to $r^{-3}$ and hence becomes important at small
distances. This term also depends on the angular momentum of a test
particle. There exists, however, a surprisingly well working
approximation to the radial potential, proposed by Paczynsky and Wiita
\cite{Paczynsky80}. This pseudo-Newtonian potential, of the form of
$V_{\rm PN}=-G\M/(r-R_{\rm S})$ where the Schwarzschild radius $R_{\rm S}=2G\M/c^2$
and $\M$ is the black hole mass, reproduces correctly positions of
marginally bound and the last stable circular orbits. The
pseudo-Newtonian potential gives efficiency factors in a good
agreement with the true solution. Since it does not depend on the
angular momentum, now the motion of a test
particle can be considered as a motion in a three-dimensional space
with the Newtonian potential replaced by the pseudo-Newtonian
one. Then, the equations of motion for the Bose-Fermi droplet moving
in the field of a fixed black hole can be put in the form which
generalizes Eqs. (\ref{eqmWD}):
\begin{eqnarray}
&& i \hbar \frac{\partial \psi_B}{\partial t} = (H^{eff}_B + V_{PN}\,m_B)\,\psi_B  \nonumber \\
&& i \hbar \frac{\partial \psi_F}{\partial t} = (H^{eff}_F + V_{PN}\,m_F)\,\psi_F 
\label{eqmWDBH}
\end{eqnarray}  \\
with $m_B$ ($m_F$) being the mass of bosonic (fermionic) atom.
See Methods for definition of other physical quantities.

\noindent
\textbf{Numerical results} \\
We solve numerically Eqs. (\ref{eqmWDBH}) in 3D by
split-operator technique \citep{Gawryluk17} for possible trajectories
of an atomic white dwarf, corresponding to closed and open orbits. We
consider the Bose-Fermi droplet consisted of $^{133}$Cs
bosonic and $^{6}$Li fermionic atoms. Such mixtures are
studied intensively experimentally
\citep{Chin17,Chin19,Weidemuller14}. To find the densities of the
Bose-Fermi droplet far away from the black hole we solve
Eqs. (\ref{eqmWDBH}) without external potential, by using the
imaginary time propagation technique \citep{Gawryluk17}. Then the
droplet is located at some distance (far away from the horizon) from
the artificial black hole and pushed perpendicularly to the radial
direction with some initial velocity. The initial position, $\vec{r}_{\rm ini}$, and
velocity, $\vec{v}_{\rm ini}$, determine the parameters of the orbit of a white dwarf. The
total energy, $E=E_{\rm k}+E_{\rm p}$, and the angular momentum, $L$,
of a white dwarf of mass $\m$ are calculated as 
$E_{\rm k}=1/2\,\m\,{\vec{v}}_{\rm ini}^2$, $E_{\rm p}=-\alpha /r_{\rm ini}$
($\alpha=G\M \m$), and $L=\m\, (\vec{r}_{\rm ini} \times \vec{v}_{\rm ini})_z$ 
with $\vec{v}_{\rm ini}$ and $\vec{r}_{\rm ini}$ being the
initial velocity of the Bose-Fermi droplet and the initial position,
respectively. The total mass of the atomic white dwarf is $\m=N_B m_B
+ N_F m_F$. The eccentricity of the orbit is given by
$\varepsilon=\sqrt{1+2 E L^2 / (\m \alpha^2)}$ and the position of the
periastron is $r_{\rm per}=p/(1+\varepsilon)$ with $1/p=\m \alpha /L^2$.

In our case we have $1042$ bosonic and $100$ fermionic atoms in the
droplet, with the interatomic forces determined by the ratio of the
scattering lengths for boson-fermion and boson-boson interactions:
$a_{BF}/a_B =-5$ (see Methods for details).
As shown in Ref. \cite{Rakshit18}, the ratio $a_{BF}/a_B$ must be smaller than $-2.8$
to be possible to form a stable ceasium-lithium droplet. The stability condition just
given is correct only for the Bose-Fermi droplet being in a free space, i.e. far away
from other objects. It slightly changes when the droplet is put in any external potential, 
in particular, the one originating from the artificial black hole. Therefore, as discussed below,
the whole process of disruption of a white dwarf begins and continues until the end of WD-BH system's life.
We consider three cases related to open and closed orbits of the WD. \\

\begin{figure}
\includegraphics[width=10.0cm]{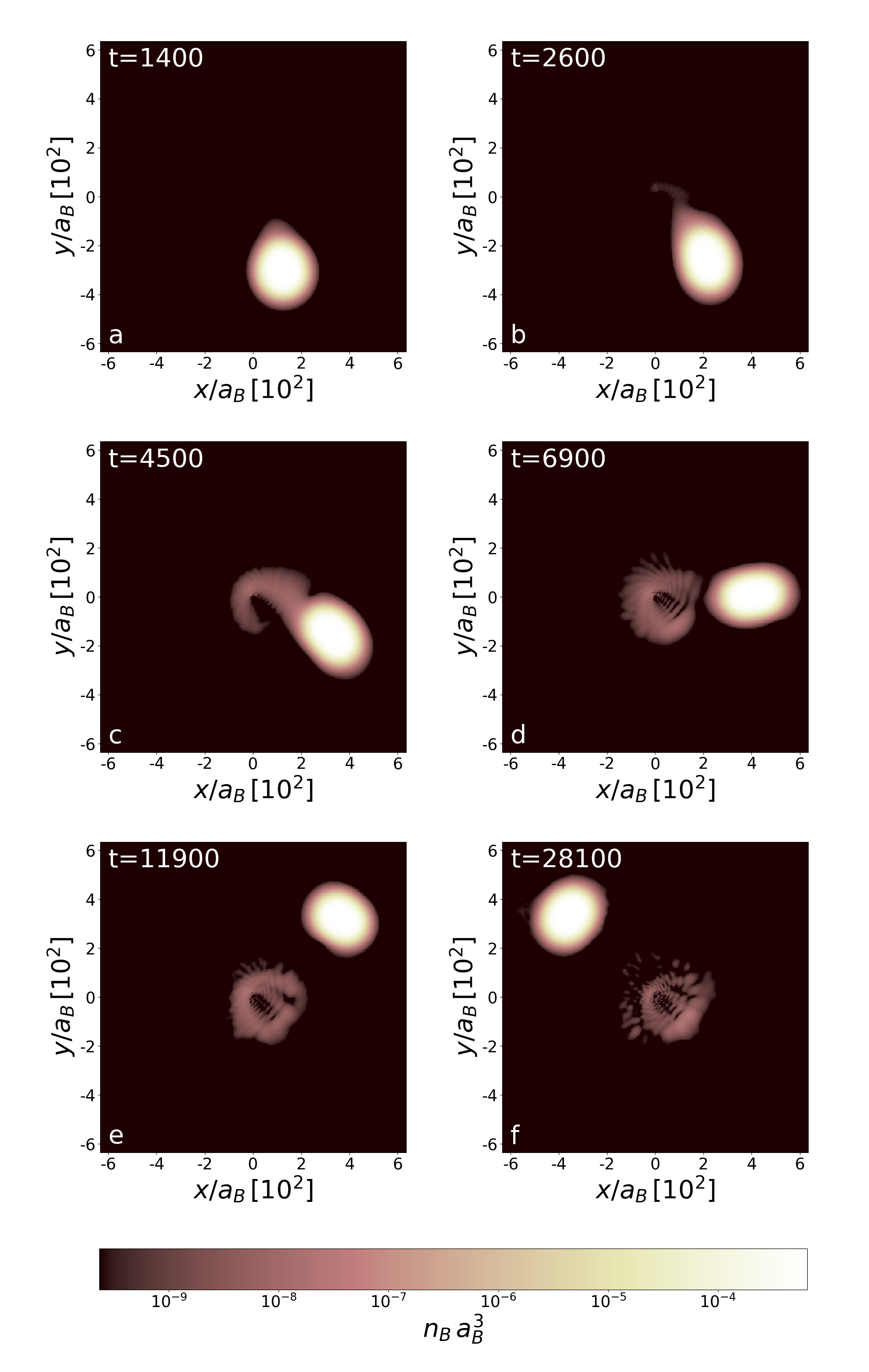} 
\caption{Density distribution (at $z=0$ plane) of bosonic component at times during (frames a-c) and after (frames e,f) the first-time periastron passage for the first considered orbit. The unit of time is $(m_B a_B^2)/\hbar$ (see Methods). Note that the distance $1$ between ticks means $10^2$ in units of $a_B$. An atomic white dwarf is stripped of about $10^{-5}$ of its mass. The stripped mass forms an accretion disk around the black hole. Frames show densities at various times during the first revolution, when mainly bosonic matter contributes to an accretion disk. The black hole is located at the center of each image and the Schwarzschild radius is small, below the unit of length $a_B$.  }  
\label{firstperiastron}
\end{figure}

\begin{figure}
\includegraphics[width=7.2cm]{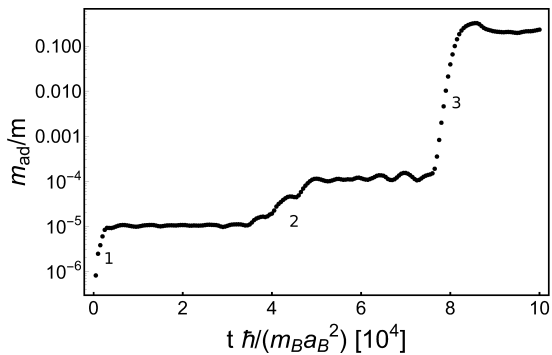}
\caption{Mass of bosonic component accumulated within the accretion disk, 
in units of $m$ ($\equiv \m$) -- the WD mass. Three periastron passages (marked by 1,2, and 3), 
equally separated in time, can be clearly identified. Each successive passage is accompanied 
by increased amount of stripped mass. At the third passage the binary ends its life, see 
Fig. \ref{thirdperiastron}. Note that the distance $1$ between ticks means $10^4$ 
in units of $m_B\,a_B^2/\hbar$.}  
\label{accretion}
\end{figure}

\begin{figure}
\includegraphics[width=10.0cm]{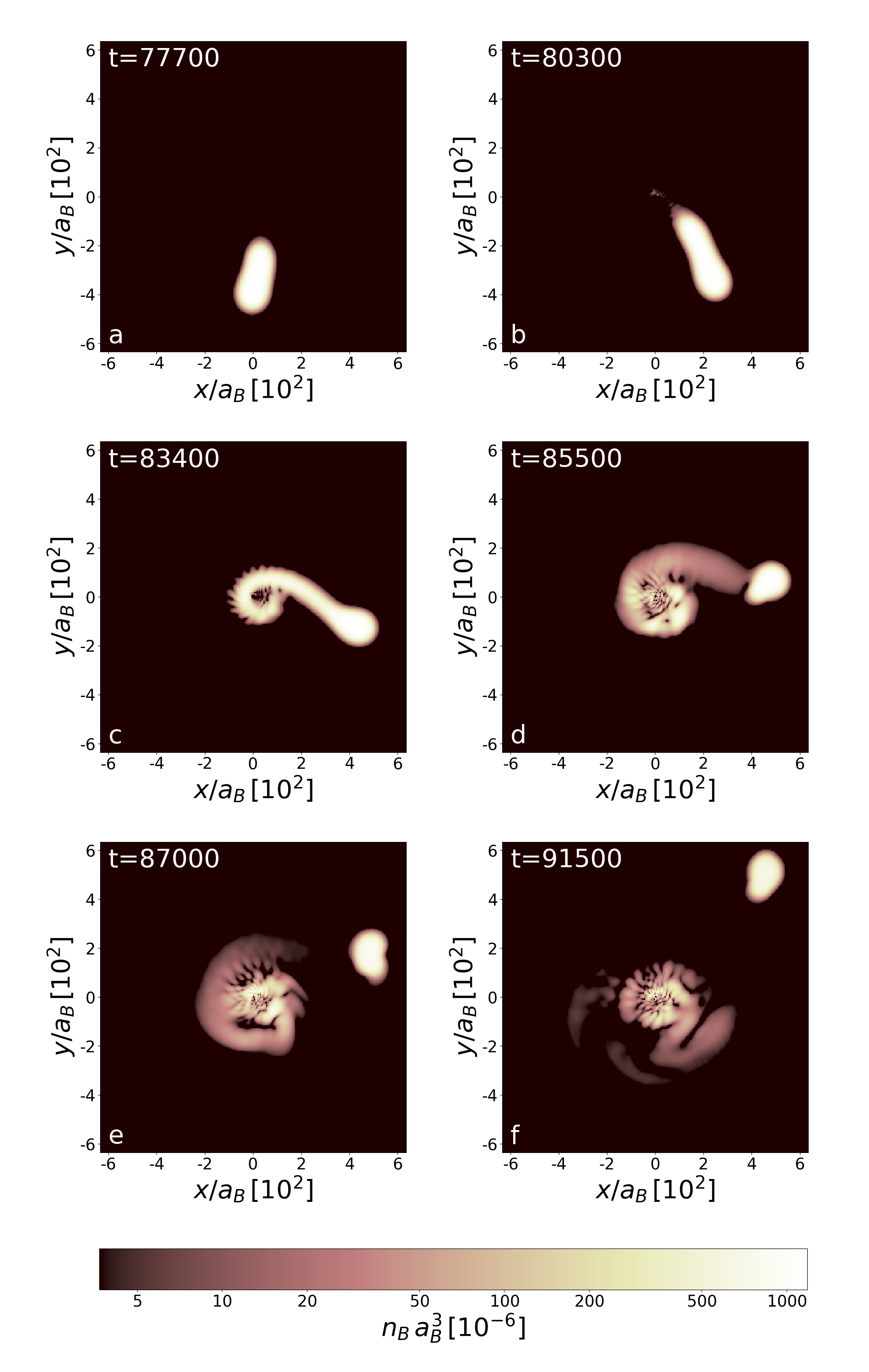} 
\caption{Density distribution (at $z=0$ plane) of bosonic component at times during (frames a-c) and after (frames e-f) the third-time periastron passage for the first considered orbit. The unit of time is $(m_B a_B^2)/\hbar$. An atomic white dwarf is stripped of about $10\%$ of its initial mass. The stripped mass (both of bosonic and fermionic type) forms a massive accretion disk. Frames show densities during the third revolution. The binary system ends its life and the white dwarf goes away of the black hole. }  
\label{thirdperiastron}
\end{figure}

\noindent
\textbf{Few periastron passages case.} 
First, we take initial dynamical parameters for the droplet such that
$E<0$, i.e. the orbit of the white dwarf is closed. We are interested
in the final stages of the evolution of black hole-white dwarf binary,
when tidal forces become damaging. It happens when the white dwarf
star itself gets larger than its Roche lobe. Only then the white dwarf
starts to loose its mass through the inner Lagrangian point L$1$. This
condition is satisfied in our simulations already for $G\M \sim 1$ and
$r_{\rm ini} \sim 100$.  All quantities are given in the code units built
of $m_B$, $a_B$, and $m_B a_B^2/\hbar$ as the units of mass (of Cesium
atom), length (a typical scattering length for atoms), and time,
respectively.  We put $G\M=1.93$ (in code units, i.e. $\hbar^2/(m_B^3\, a_B)$), 
an estimation of the mass of the black hole is given at the end of the subsection.  
For $\vec{\upsilon}_{\rm ini}=(0.09,0,0)\, \hbar/(m_B\, a_B)$ 
(white dwarf released along $x$ axis) and $r_{\rm ini}=320$ one has
$r_{\rm per}=320$ and the white dwarf is initially at the periastron.
The simulations show that the white dwarf circulates the
black hole only a few times. It is significantly stripped off the mass
after each periastron passage, many orders of magnitude stronger than
the estimation given in \cite{Shen18}. Already at the first observed
passage the white dwarf looses $10^{-5}$ of its mass, see
Figs. \ref{firstperiastron} and~\ref{accretion}. The mass loss at
the third passage is extremely large (Fig. \ref{accretion}) and
results in the death of a white dwarf-black hole system
(Fig. \ref{thirdperiastron}) -- the white dwarf is running away.

Fig. \ref{firstperiastron} illustrates the way the white dwarf is
stripped of its mass. The pipe connecting the black hole and the white
dwarf is open extremely fast when the star is passing through the
periastron for the first time, frame \ref{firstperiastron}(b). The
charged mass falling onto the black hole becomes a source of powerful
radiation. The time corresponding to the opening of a pipe is simply
interpreted as a rise time of a signal, an X-ray burst as argued in
\cite{Shen18}, detected by the observer. The mass is stripped off a
white dwarf during almost a quarter of the circulation period, see
frames \ref{firstperiastron}(c) and \ref{firstperiastron}(d), and the
accretion disk is created. Since the radiation is mostly emitted in
the direction perpendicular to the pipe's direction, its amount
reaching the observer decreases. Hence, the signal decays and the
falling time is estimated to be about $50$ times longer than the rise
time, the ratio similar to that reported in \cite{Irwin16} for NGC
4697 and NGC 4636.  Finally, the pipe is broken (frame
\ref{firstperiastron}(d)) and the radiation is seized at the
background level until the white dwarf, while circulating the black
hole (frames \ref{firstperiastron}(e,f)), enters periastron region
again (Fig. \ref{accretion}).

Figs. \ref{firstperiastron}(d,e,f) show the accretion disk appearing
around the black hole after the white dwarf is stripped off its mass
for the first periastron passage. The size of the accretion disk,
i.e. the orbiting material gravitationally bound to the black hole, is
about one half of the distance between components. Indeed, the
existence of flat parts in the curve in Fig. \ref{accretion}, showing
the accreted mass $m_{\rm ad}$ as a function of time, proves that the
mass of the accretion disk is essentially confined in a disk of half
of the black hole-white dwarf separation. 

Eventually, after a huge loss of mass during the third periastron
passage, the white dwarf is expelled out of the neighbourhood of the
black hole, see Fig. \ref{thirdperiastron}. The binary system ends its
life. The matter remained in the accretion disk becomes fragmented due
to modulational instability \citep{Kivshar98} -- a nonlinear effect,
both classical and quantum, closely connected to the existence of
solitary waves, already observed for Bose-Einstein condensates
\citep{Nguyen17,Everitt17}.

\begin{figure}
\includegraphics[width=8.0cm]{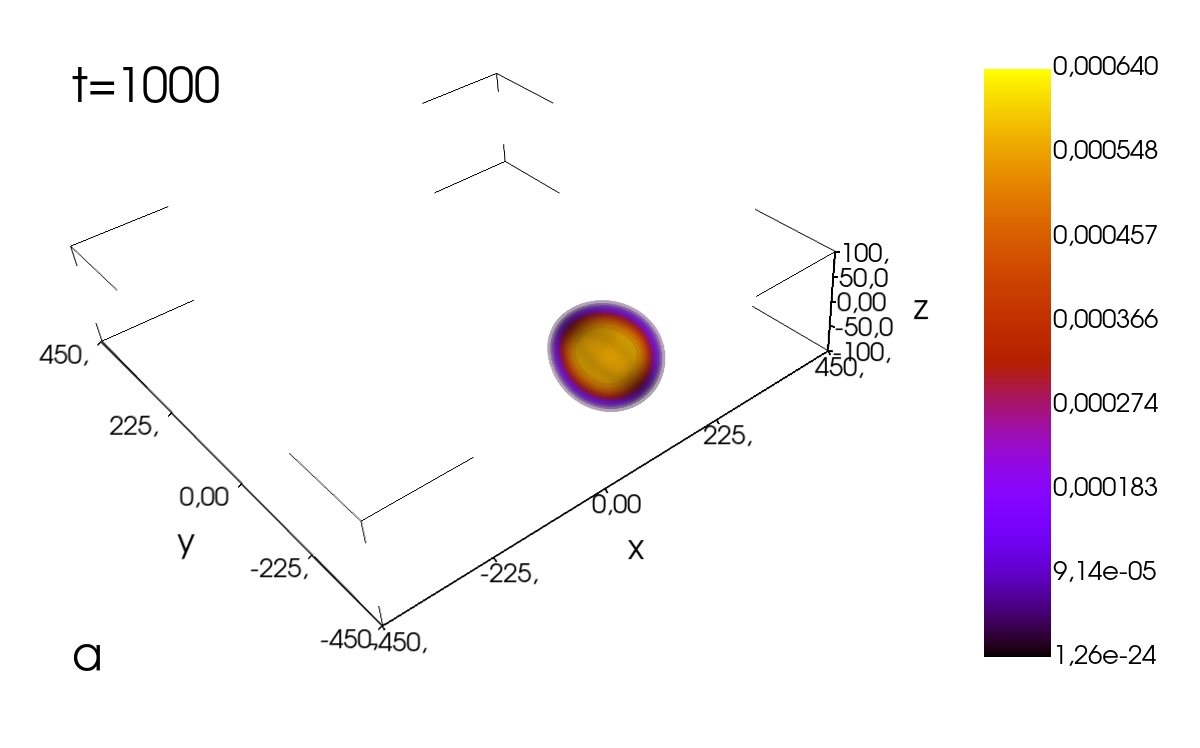}  \includegraphics[width=8.0cm]{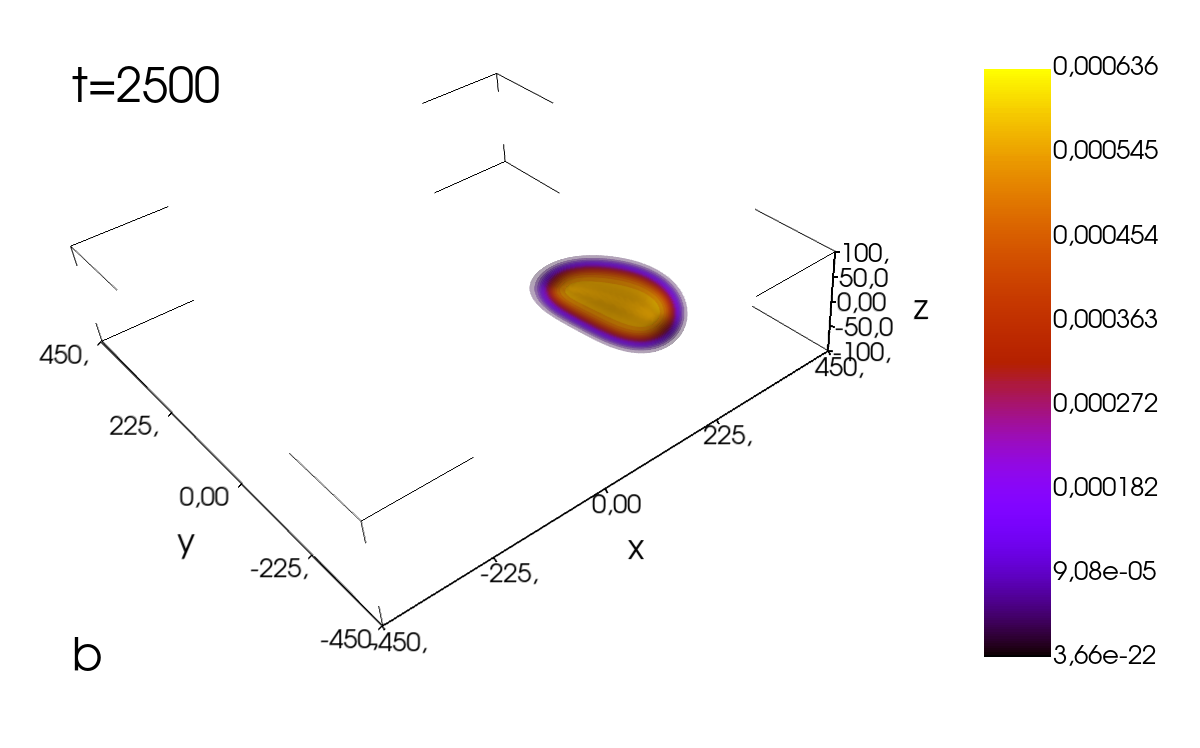} \\   
\vspace{0.4cm}
\includegraphics[width=8.0cm]{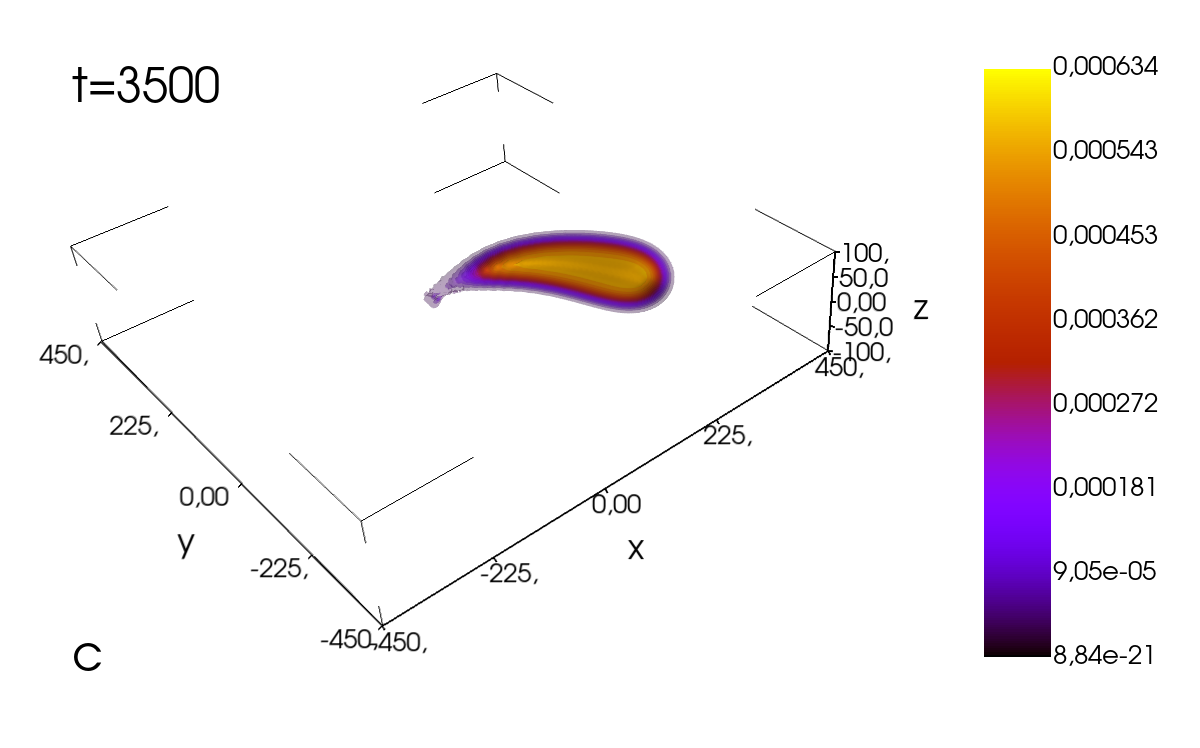}  \includegraphics[width=8.0cm]{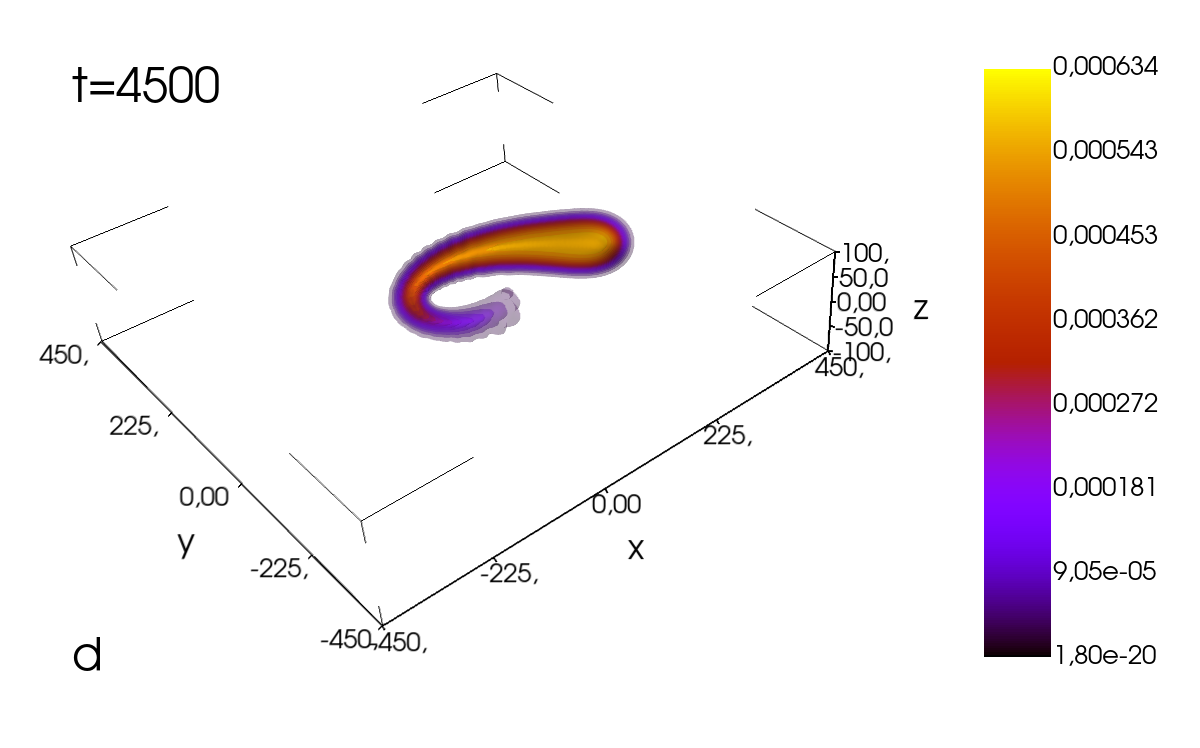} \\   
\vspace{0.4cm}
\includegraphics[width=8.0cm]{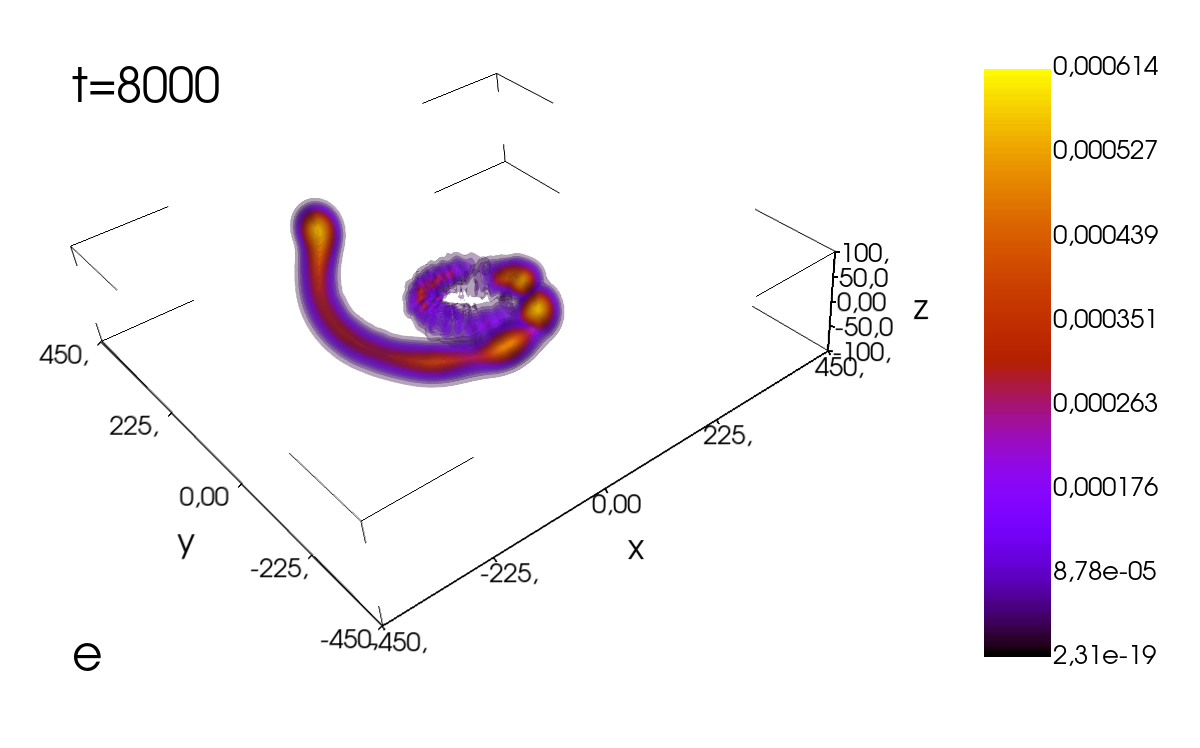}  \includegraphics[width=8.0cm]{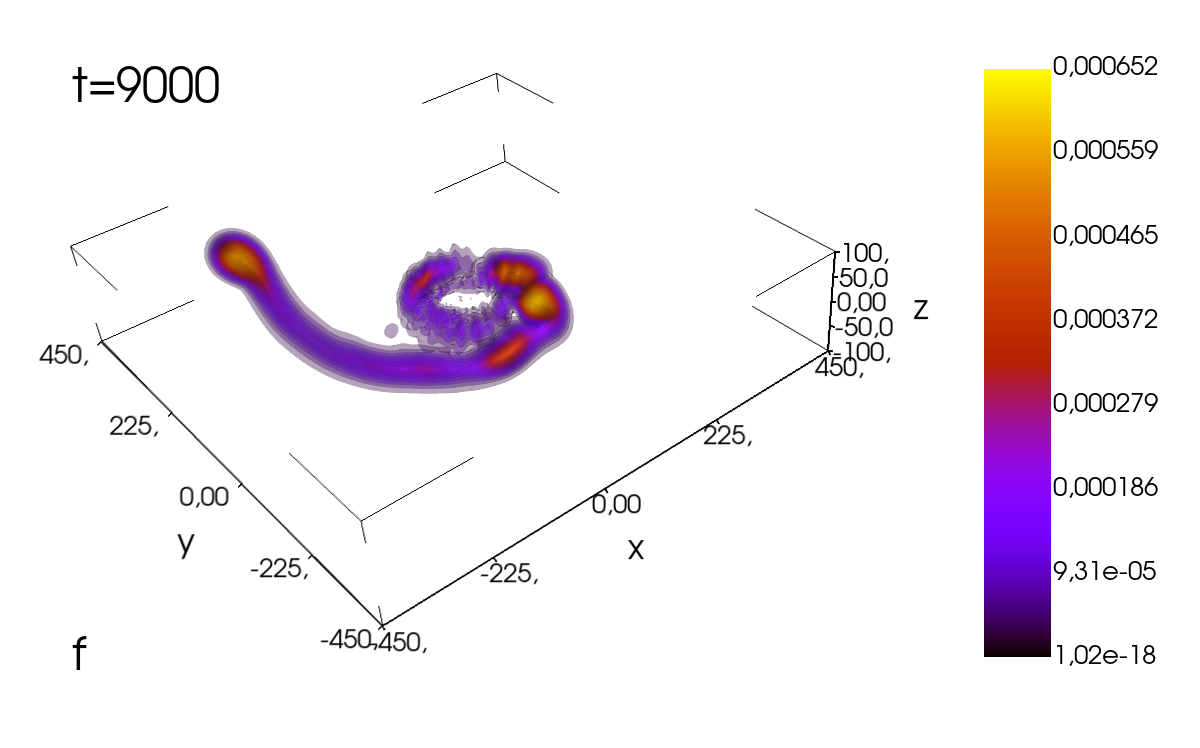}
\caption{Atomic white dwarf in the field of the black hole, located at the center of each box.
The white dwarf is continuously stripped off its mass and an accretion disk is formed. At the final stage, 
the fragmentation of the orbiting material is clearly visible. Also, giant quantized vortices 
are formed in the bosonic accretion disk (see Fig. \ref{vortices} for closer inspection). 
The units of distance, time, and density (color bars) are $a_B$, $(m_B a_B^2)/\hbar$, 
and $a_B^{-3}$, respectively. }
\label{secondcase}
\end{figure}

Although only bosonic component densities are shown in Figs. \ref{firstperiastron} and \ref{thirdperiastron}, it is true that fermionic matter contributes to the accretion disk as well. Only during the first revolution the accretion disk is mainly formed from bosons. This is because the external field of a black hole changes the stability condition for the white dwarf and relative number of bosons and fermions in the white dwarf must change. In our case, an excess bosonic matter falls on a black hole during the first revolution. However, later on, when the action of a black hole on a white dwarf gets stronger, the white dwarf is stripped equally of bosonic and fermionic matter (Fig. \ref{thirdperiastron} and also Fig. \ref{secondcase}).

An estimation of the mass of the black hole can be done based on the
assumptions that at the periastron the white dwarf overfills its Roche
lobe. This is, of course, the case since we observe the transfer of
mass from the white dwarf to the black hole's Roche lobe. According to
the formula given in \cite{Sepinsky07}, which is an extension of
earlier works \citep{Paczynski71,Eggleton83} to the case of eccentric
orbits and nonsynchronous motion, the radius of the Roche lobe for the
white dwarf at the periastron becomes $R_{\rm L}=r_{\rm per}\times 0.49\,
(\m/\M)^{1/3}$, and it is assumed that the ratio $\m/\M$ is
small. The mass of the black hole is estimated from the inequality
$R_{\rm L}\lesssim r_{\rm wd}$, where $r_{\rm wd}$ is the radius of the white
dwarf. This gives the lower limit for the mass of the black hole
$\M/\m \gtrsim (r_{\rm per}/(2\, r_{\rm wd}))^3$, equal about $8$ in our case.
For this lower limit, the dimensionless penetration parameter $\beta\
(\equiv r_{\rm t}/r_{\rm per})$ equals to $0.5$, where $r_{\rm t}$ is the tidal radius. \\

\begin{figure}
\includegraphics[width=9.0cm]{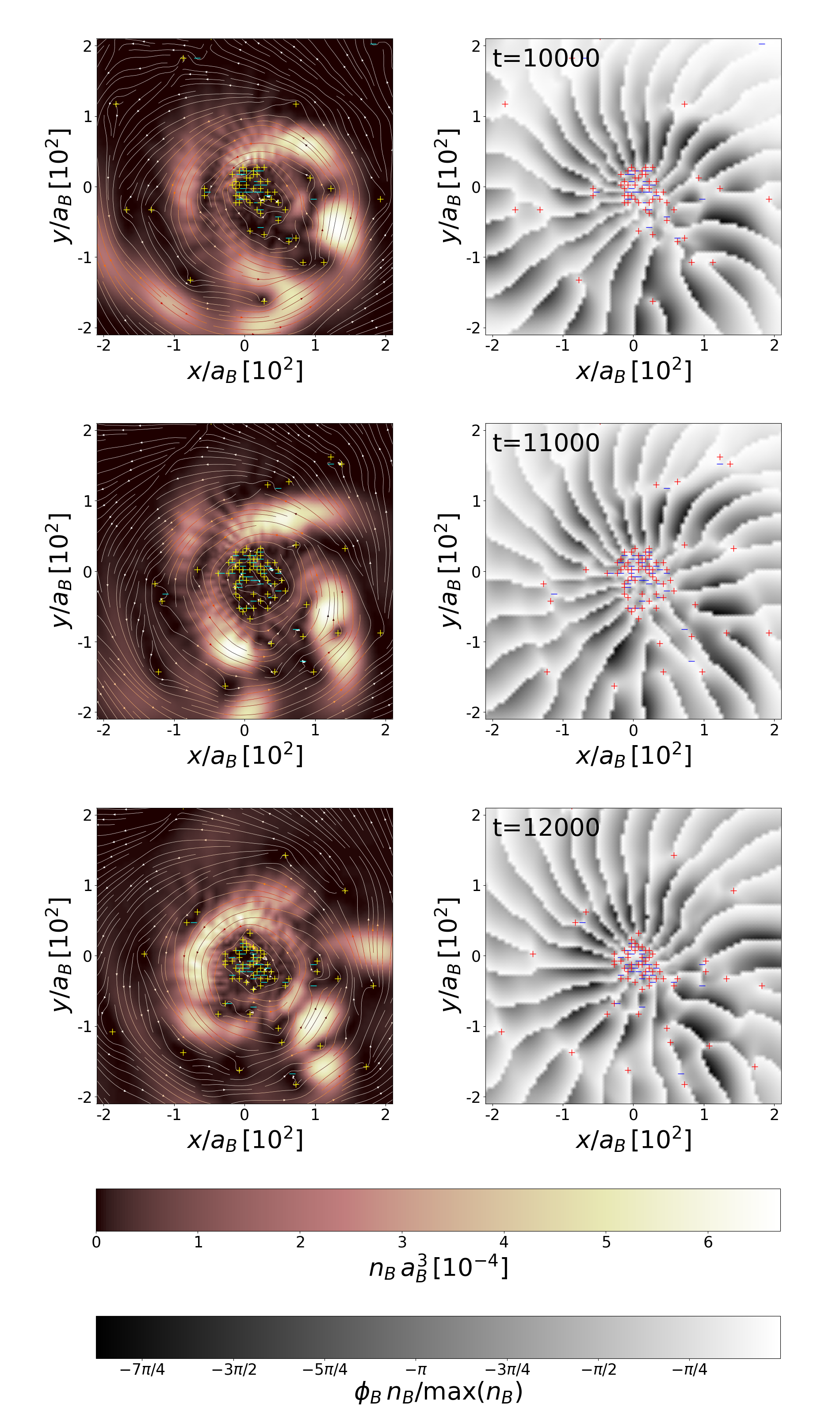}
\caption{Density with streamlines (left frames) and the phase (right frames)  
of the bosonic accretion disk at different times, increasing from top to bottom. 
The unit of time is $(m_B a_B^2)/\hbar$. The black hole is in the (0,0) position.
The quantized vortices are visible as plus and minus signs present within the matter 
spiraling around the black hole, at the inner and outer edges of the matter spout, 
as well as in the region of low density. }  
\label{vortices}
\end{figure}

\noindent
\textbf{One periastron passage for an elliptical orbit.} 
For the second case, we consider, the system's energy is still
negative, $E<0$, but the attraction between the black hole and the
white dwarf is now stronger ($G\M=3.87$). Then, for $\vec{\upsilon}_{\rm ini}=(0.09,0,0)$
and $r_{\rm ini}=320$, the periastron equals $r_{\rm per}=160$. 
In this case the parameter $\beta = 0.63$.
The white dwarf is continuously stripped off its matter
and an accretion disk appears around the black hole,
Fig.~\ref{secondcase}. The orbiting material becomes fragmented due to
nonlinear effects and giant vortices are formed in the bosonic
component (see Fig.~\ref{vortices}). Creation of giant vortices might
be responsible for another strong bursts of radiation since the radial
acceleration of particles moving within the quantized vortex is
$\propto 1/r^3$ ($r$ is a distance from the vortex core) which is
stronger than the acceleration in the field of Newtonian potential
($\propto 1/r^2$).  Eventually, the white dwarf goes away from the
black hole ending the life of the binary system.

Fig. \ref{vortices} clearly shows that quantized vortices of a single
charge with both signs of vortivity are nucleated in a large number in
the accretion disk in bosonic component while the matter is falling
onto the black hole. They are formed at the inner and outer edges of
the matter spout and some of them move into the region of high
density. The size of the vortices cores is of the order of the healing
length $\xi=\hbar /\sqrt{g_B n_B}$ and for the vortices located in the
high density region it is as large as $15 \%$ of the diameter of the
white dwarf. Most of the vortices are created in the region of low
density, in the neighbourhood of the black hole. Since the size of the
vortex core increases with lowering density, vortices must overlap.
Their motion is highly irregular, spatially and temporally
disordered. Their number changes significantly, for instance it goes
up by about $30 \%$ while going from the upper to the middle frame in
Fig. \ref{vortices}. Then it falls down for the lower frame in
Fig. \ref{vortices}. The superfluid Reynold number $Re_{\rm s}=m_B \xi 
\upsilon /2\pi\hbar$, where $\upsilon$ is the flow velocity, takes values in a wide
range, of the order from $10^{-3}$ up to $10^2$.  All of these suggest
the onset of a quantum turbulence \citep{Bagnato16} in the accretion
disk.  Since the accretion disk is an oblate object, one should think
here rather about two-dimensional quantum turbulence
\citep{Bradley12,Neely13}. 
Vortices, as topological objects, are indeed very stable structures. As demonstrated
in numerous experiments on cold atoms, both condensed bosons and superfluid fermions 
\cite{Ketterle01,Ketterle05}, they survive during the system's expansion after the trapping 
potential is turned off. They survive and scale up together with the whole system.
As robust objects, we expect that vortices should remain present in the system while 
amplifying the system from micrometer-size to astronomical ones. \\

\begin{figure}
\includegraphics[width=10.0cm]{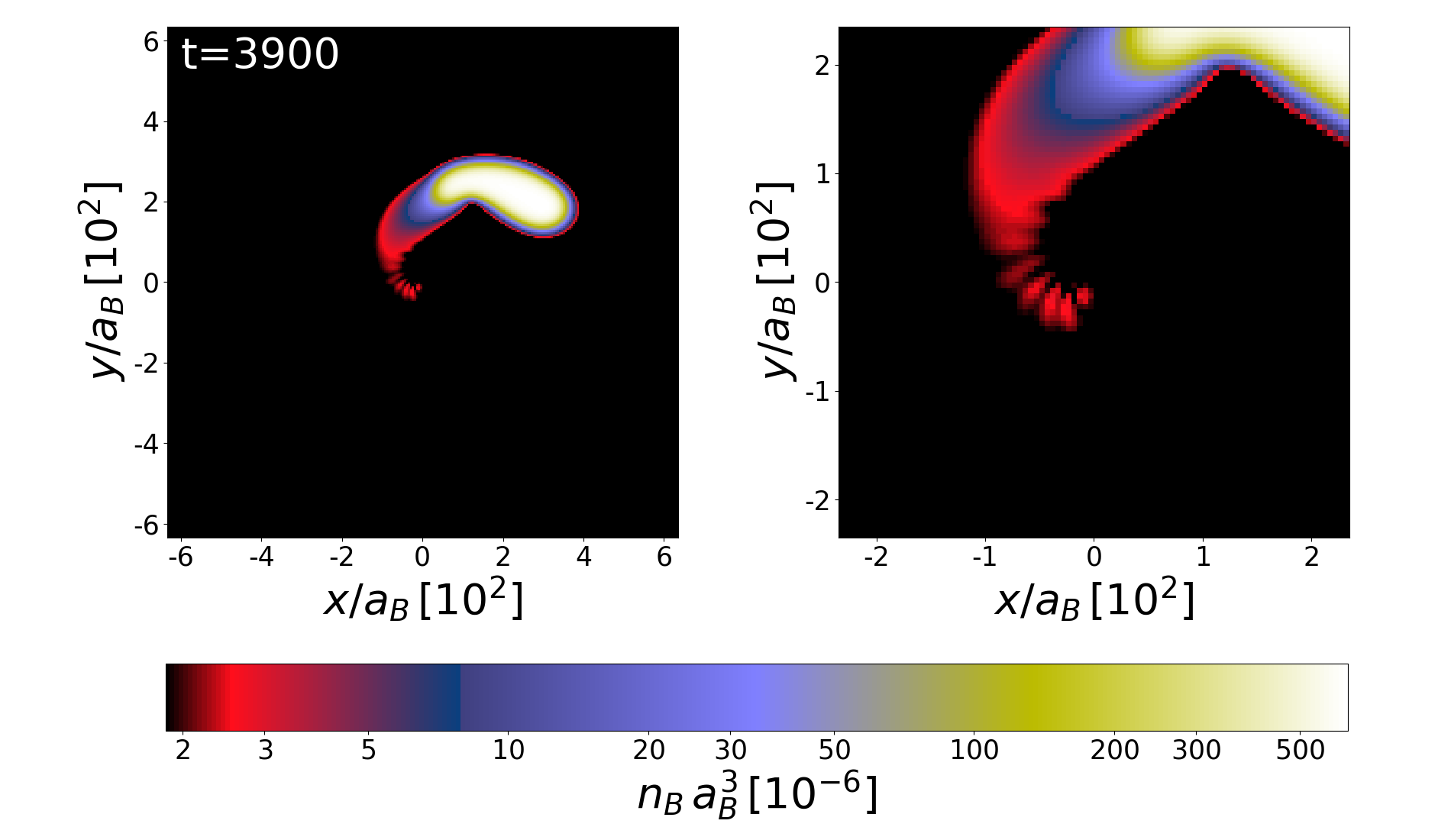}
\caption{Grains of matter falling onto the black hole while the white dwarf is 
passing aside, along the hyperbolic orbit. A series of falling matter blocks 
(visible in red color) could be responsible for radiation of a train of 
ultraluminous flares. The right frame magnifies the central part of the left 
image. The unit of time is $(m_B a_B^2)/\hbar$.}
\label{case3}
\end{figure}

\noindent
\textbf{Hyperbolic orbit case.} 
Finally, we consider the case when the total energy is positive,
$E>0$, with $G\M=1.93$, $r_{\rm ini}=320$, and $\vec{\upsilon}_{\rm ini}=(0.1,0.1,0)$. 
It corresponds to $\beta = 0.84$. As the
white dwarf passes the black hole, the part of it is expelled towards
the black hole, see Fig.~\ref{case3}. Once again, due to nonlinear effects one
can observe, in fact, a sequence of blocks of matter falling onto the
black hole. Consequently, grains of falling matter are expected to emit trains 
of ultraluminous flares.  \\

\noindent
\textbf{Discussion} \\
We would like to emphasize the presence of two new phenomena,
  uncovered by quantum hydrodynamic simulations, which play an
  important role in the physics of the accretion disks,
  i.e. fragmentation of falling matter and creation of quantized
  vortices. Both these exceptional phenomena are a manifestation of
  coherence of bosonic component inside the accretion disk.  In order
  to find astrophysical consequences of them we need to rescale the results of our
  numerical simulations to astronomical objects.

  This can be easily done by
  multiplying the distance, the mass and the numerically normalised Planck 
  constant by $a$, $b$, and $c$, respectively (see Methods for
  details). Then, consequently, the time and the energy are scaled as
  $b\,a^2/c$ and $c^2/(b\,a^2)$ and the hydrodynamic equations
  (Eqs. (\ref{hydroFB}) in the Methods), basic for
  our modelling, do not change. Then by choosing appropriately large
  $a$ and $b$ one can increase the size, the mass as well as the mass
  density (since it is scaled as $b/a^3$) of the object which we
  model. All contributing energy ingredients are, of course, scaled in
  the same way. It means that the attractive interaction energy
  between droplet's particles gets large on the same footing as the
  kinetic energy of fermions, maintaining the stability of
  studied system.

  It is worth to mention that our model is capable to
  explain the origin of radiation coming out of the black hole-white
  dwarf binary system. It turns out that the equilibrium condition for
  the white dwarf in the presence of additional potential originating
  from the black hole is different from the one corresponding to the
  case when the white dwarf stays far away from any sources of
  disturbance. Different here means other number of atoms in the
  system which remains still bound. Therefor, in the presence of the
  black hole some part of bosonic or fermionic component has to be
  expelled from the white dwarf.  Then the system, which was
  originally neutral, becomes charged.  Consequently, grains of
  falling charged matter are forced to emit radiation. 

Perhaps, a recent archival X-ray data search of nearby galaxies
uncovering two sources of ultraluminous flares \citep{Irwin16} could
be explained by our simulations.  One of those sources flared once
with estimated peak luminosity of $9 \times 10^{40}$erg/s, the second
one flared five times and its intensity was about ten times weaker.
All these X-ray bursts have similar rise time, which is less than one
minute and the decay time of about an hour. Together with other
flaring X-ray source found earlier in NGC 4697 \citep{Sivakoff05} they 
might constitute a new astrophysical phenomena -- a new type of fast
transients. So far known non-recursive and recursive transient
phenomena should be excluded as an explanation of reported flare
sources because of one of the following reasons: too low luminosity,
too long duration, or inappropriate location of the source. An
explanation of observed X-ray bursts as originating from a tidal
stripping of a white dwarf circulating an intermediate-mass black hole
has just been proposed \citep{Shen18}. Our simulations seem to support
this explanation.
In 2011 \cite{Burrows11} reported the observation of a bright
X-ray flare from the extragalactic transient Swift
J164449.3-573451. This event, supported by the optical, infrared, and
radio observations \citep{Zauderer11,Levan11} has been related to the
disruption of a star by a massive black hole located at the center of
a distant galaxy. What was unusual in measured signal was its internal
structure. In fact, an irregular sequence of brief flares was
detected. \cite{Krolik11} argued that the multiple recurring hard
X-ray bursts emitted by Swift J164449.3-573451 object could originate
in the disruption of a white dwarf by an intermediate mass black hole.
Again, our simulations seem to support this point of view (see
Fig. \ref{case3}) -- the presence of flare trains, caused by 
the fragmentation of the accreted matter.

Scaling up the simulation results to an astronomical system (see Methods) indicates 
that there occurs a sequence of flares with increasing peak luminosity
($L > 10^{40}$ erg s$^{-1}$) for each passage through periastron.
The final passage of a tidally disrupted white dwarf causes a burst, supposedly the gamma-ray burst (GRB)
with the peak luminosity of the order of $10^{52}$ erg s$^{-1}$. Its life span is $\sim 1$ s
in the case of the stellar black hole.
Two earlier passages (see Fig. \ref{accretion}) result in bursts of peak luminosities of $10^{49}$ erg s$^{-1}$
and of $10^{48}$ erg s$^{-1}$, respectively. These durations and luminosities are typical for short GRBs.


The formation of vortices is the natural explanation for the presence
of the flicker noise. A typical variability manifested in accretion disks
of the cataclysmic variables, the X-ray binaries with neutron stars or black holes,  
and active galactic nuclei is of the stochastic nature
\citep[e.g.][]{Mushotzky93,Markowitz04,Niko04,Gierl08,Miniutti09,
  Andrae13,Vagnetti16,Balman19}.  The broadband variability is often
quantified based on the power spectral density (PSD) technique where
on the low frequencies it shows a flat spectrum ($\propto \nu^0$) and
on the higher frequencies PSD follows the flicker noise or the red
noise (PSD $\propto \nu^{\alpha}$, $\alpha = -1$,$-2$, respectively).
The underlying process is still unknown. Possible explanations for
those phenomena include accretion disk instabilities, magnetic flares
above the accretion disk, viscous radial inflow, or changes in the accretion
rate \citep[e.g.][]{Kawaguchi98,Goosmann06,Sobolewska11,Marin2017}.
Phenomenological approaches are also used. One of them is the hot-spot
model where the signal is generated by an ensemble of spots randomly
created on the accretion disk surface
\citep{Czerny04,Trzesniewski11,Pechacek13}.  Our simulations can
explain production of hot-spots. We observe that the quantized vortices appear in the accretion disc.
The thermal component, as opposed to the condensed matter, does
not participate in quantum circulation  and has only Keplerian motion around the black hole.
The number, locations and sizes of quantized vortices change in time.  
Since giant vortices are responsible for strong bursts of radiation, they naturally can be regarded as hot-spots.

\begin{figure}
\includegraphics[width=10.0cm]{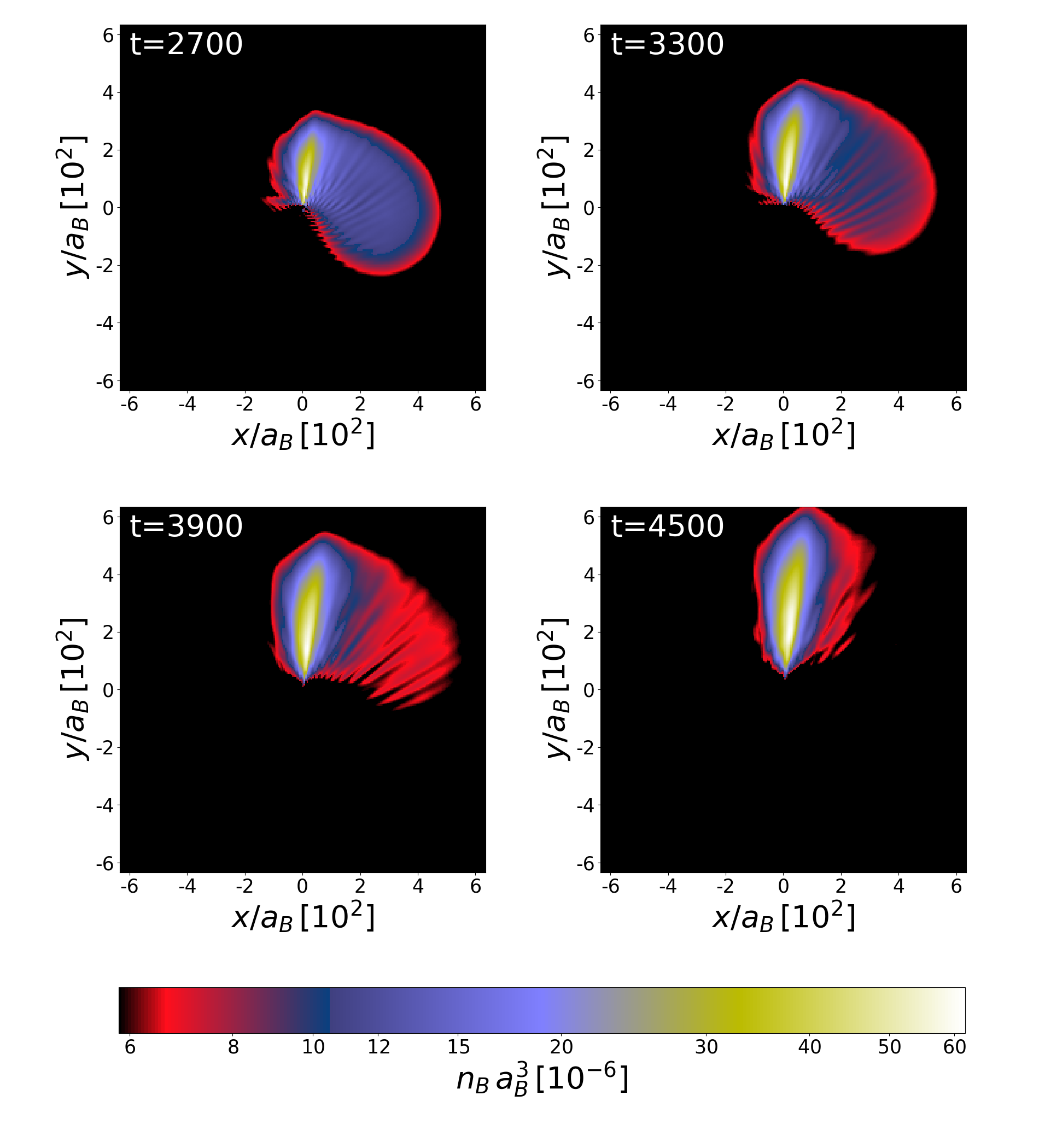}
\caption{Densities of the white dwarf at different times (frames a-d)
  while orbiting the black hole along the hyperbolic orbit as in
  Fig. \ref{case3}. Here, the Eqs. (\ref{eqmWDBH}) are solved without
  quantum corrections. The system is unstable and its disruption looks
  differently than in Fig. \ref{case3}. The bulk of the white dwarf
  flows around the black hole rather than it drops on it in pieces,
  resembling a bit results obtained within the SPH simulations
  \citep{Kawana18}. The unit of time is $(m_B a_B^2)/\hbar$. } 
\label{noqcorr}
\end{figure}


  Further comments regarding formation of the accretion disk can be done. 
  A violent change of the density in the accretion disk,
  Fig. \ref{vortices}, might suggest that the significant part of the
  white dwarf matter was reheated while falling onto the black
  hole. We have estimated, by using the classical field approximation
  \citep{review}, an amount of bosonic matter dragged out of the
  condensate during the fall. It clearly turns out that two stages of
  the dynamics can be distinguished. First, during the formation of
  the accretion disk (frames c-f in Fig. \ref{secondcase}) the
  nonlinear phenomena take advantage leading to fragmentation of the
  matter. But when the white dwarf is gone and the accretion disk is
  no more replenished with matter, the second stage begins. We
  observe slow thermalization of the accretion disk. After additional
  time equivalent to that covering frames a-f in
  Fig. \ref{secondcase}, the noncondensed fraction in bosonic matter
  increases up to $20\%$. Moreover, it should be noted, that in cases
  of strong tidal interactions (i.e. $r_{\rm per} \lesssim 0.05 r_{\rm t}$) 
the tidal compression could
  additionally trigger explosive thermonuclear reactions
  \citep{Kawana18,Anninos18}.

Now, let us focus on the stability of Bose-Fermi droplets, we use
to model a white dwarf, and the consequences of that. The Bose-Fermi
droplets are intrinsically stable.  They possess a well defined
surface. To trace the significance of the stability of the system, we
repeated simulations but this time quantum corrections (the
Lee-Huang-Yang correction for bosons and the Viverit-Giorgini one for
boson-fermion interaction, see the Methods) have been
omitted. The results are shown in Fig. \ref{noqcorr}. The white dwarf
bulk is not falling onto the black hole now, it is rather flowing
around the black hole and expanding simultaneously. No fragmentation
occurs, hence the formation of any bursts of radiation is closed. Then
the stability seems to be the necessary ingredient in the modeling of
dynamics of the white dwarf. For example, the accretion disk is
created only provided the quantum corrections for the white dwarf are
included to oppose to what other non-quantized approaches predict. 

Finally, we would like to discuss some further possible extentions
to simulations we have presented. As argued in the text, ceasium atoms play
the role of Helium ions in real WD. Fermionic lithium is the origin of the Fermi 
pressure, lithium atoms then do what electrons do in real WD. Expansion due to 
the Fermi pressure is stopped by attraction between all the particles in the 
system -- this is the origin of stability of WD. This stability is fundamental 
for true WD. Indeed, constituents of a real WD are charged particles but, in fact, 
coulombic interactions do not play any role in stabilizing WD. Different mass ratio 
of Cs/Li with respect to Helium ion/electron might change the response of WD to external 
field. We believe that only quantitatively. Properties of BH-WD binary analyzed from this 
perspective could be the subject of further studies, in which artificial atoms 
(i.e. with arbitrary mass ratio) would be used to form a Bose-Fermi droplet. \\


\noindent
\textbf{Conclusions} \\
In summary, we have studied dynamics of a cold white dwarf star in the
field of a black hole. As a model of a white dwarf star we consider a
zero temperature droplet of attractively interacting degenerate atomic
bosons and spin-polarized atomic fermions. Our quantum hydrodynamics
based simulations reveal unexpected behavior of the black hole-white
dwarf binary, particularly at the end of its existence: 

\begin{itemize}
\item Our calculations show the fragmentation of the falling matter.
  It supports the hypothesis that the binary system
  could be responsible for the recently reported ultraluminous X-ray
  bursts. We predict the possibility of trains of such flares as well.

\item Giant quantized vortices appear in the accretion disk. They
      could constitute another source of radiation, in particular in a
      connection with the quantum turbulence.

\item The accretion disk, if we take into account quantum corrections
      for the white dwarf, is created to oppose to what other
      non-quantized approaches predict.
      
\item The gamma-ray burst (GRB) could be a product of the final passage of a white dwarf 
near a black hole just before destruction of the WD-BH system.      
\end{itemize}

All of these phenomena happen because nonlinear
and quantum effects manifest on the same footing while the white dwarf
meets the black hole. \\

\noindent
\textbf{Corresponding author} \\
Marek Nikolajuk: m.nikolajuk@uwb.edu.pl \\

\noindent
\textbf{Methods} \\ 
\textbf{Quantum hydrodynamic equations for white dwarfs.}
The quantum hydrodynamic equations for the white dwarf modeled as the
Bose-Fermi quantum droplet are given by
\begin{eqnarray}
\frac{\partial\, n_{F}}{\partial t} &=& -\nabla \cdot \left(n_F\, {\bf{v}}_{\!F}    \right),   
\nonumber \\
m_F\frac{\partial\, {\bf{v}}_{\!F}}{\partial t} &=& -\nabla\left(\frac{\delta T}{\delta 
n_F}+\frac{m_F}{2} {\bf{v}}_{\!F}^2 + \frac{\delta E_{BF}}{\delta n_F}  \right) \nonumber \\
\frac{\partial\, n_B}{\partial t} &=& - \nabla \cdot \left( n_B {\bf{v}}_{\!B} \right)  \nonumber \\
m_B \frac{\partial\, {\bf{v}}_{\!B}}{\partial t} &=& - \nabla \left( \frac{\delta E_B}{\delta n_B} + 
\frac{m_B}{2} {\bf{v}}_{\!B}^2 + V_q + \frac{\delta E_{BF}}{\delta n_B} \right)  \,,   
\label{hydroFB}
\end{eqnarray}
where $n_F({\bf r},t)$ and $n_B({\bf r},t)$ are the densities of
fermionic and bosonic fluids, respectively and ${\bf{v}}_{\!F}({\bf r},t)$ 
and ${\bf{v}}_{\!B}({\bf r},t)$ are the corresponding
velocity fields. These equations can be derived based on quantum
kinetic equations for reduced density matrices
\citep{Frohlich,Wong,MarchDeb}. $T$ is the intrinsic
kinetic energy of an ideal Fermi gas and is calculated including
lowest order gradient correction only
\citep{Weizsacker,Kirznits,Oliver}

\begin{eqnarray}
T = \int d{\bf r}\, \left( \kappa_k\,n_F^{5/3}  -\xi\, \frac{\hbar^2}{8m_F} \frac{(\nabla n_F)^2}{n_F} \right)
\label{T3D}
\end{eqnarray}
with $\kappa_k = (3/10)\,(6\pi^2)^{2/3}\,\hbar^2/m_F$ and $\xi=1/9$.
There is an additional term in the second equation in (\ref{hydroFB})
related to the other, bosonic component of the droplet. Bosons and
fermions interact and in the simplest, mean-field, approximation the
interaction energy is $E_{BF}^{mf} = \int d{\bf r}\, g_{BF}\, n_B({\bf
  r}) n_F({\bf r})$. To stabilize the Bose-Fermi droplet \citep{Rakshit18} 
the quantum correction due to quantum
fluctuations is necessary $E_{BF}^{qc} = C_{BF}\int d{\bf r}\, n_B\,
n_F^{4/3} A(w,\gamma)$, where $w=m_B/m_F$ and $\gamma=2w (g_B
n_B/\varepsilon_F)$ are dimensionless parameters, $C_{BF}=(6
\pi^2)^{2/3} \hbar^2 a_{BF}^2 / 2 m_F$, and the function $A(w,\gamma)$
is given in a form of integral \citep{Giorgini02}
\begin{eqnarray}
A(w,\gamma) = \frac{2(1+w)}{3w}\left(\frac{6}{\pi}\right)^{2/3}\int^{\infty}_0 {\rm d}k 
\int^{+1}_{-1} {\rm d}{\Omega}
\left[ 1 -\frac{3k^2(1+w)}{\sqrt{k^2+\gamma}}
\int^{1}_0{\rm d}q\, q^2 \frac{1-\Theta(1-\sqrt{q^2+k^2+2kq\Omega})}{\sqrt{k^2+\gamma}+wk+2qw\Omega}  
\right], \nonumber\\
\label{A}
\end{eqnarray}
with $\Theta()$ being the step theta-function. Then the total
boson-fermion interaction energy is $E_{BF} = E_{BF}^{mf}+E_{BF}^{qc}$.
$V_q=-\hbar^2/(2 m_B)\, (\nabla^2\sqrt{n_B}) /\sqrt{n_B}$ is related
to the bosonic quantum pressure \citep{Madelung}. The first term in
the Euler-like equation for bosons is due to interaction between
bosons, $E_B = g_B n_B^2/2+E_B^{LHY}$, including the famous
Lee-Huang-Yang correction $E_B^{LHY} = C_{LHY} \int d{\bf r}\,
n_B^{5/2}$ with $C_{LHY}=64/(15\sqrt{\pi})\,g_B\, a_B^{3/2}$
\citep{LHY57}.  $g_B$ and $g_{BF}$ appearing in the above energy
expressions are coupling constants for contact interactions between
atoms \citep{PitaevskiiStringari}, with $g_B = 4\pi \hbar^2 a_B/m_B$
and $g_{BF} = 2\pi \hbar^2 a_{BF}/\mu$, where $a_B$ ($a_{BF}$) is the
scattering length corresponding to the boson-boson (boson-fermion)
interaction and $\mu=m_B\, m_F/(m_B+m_F)$ is the reduced mass.

Eqs.  (\ref{hydroFB}) constitute of continuity equations
for bosonic and fermionic fluids and of equations governing the
motions of fluid elements under the presence of forces originating
from the quantum pressure and inter- and intra-species
interactions. Equations describing bosons are just the hydrodynamic
representation of the Gross-Pitaevskii equation
\citep{PitaevskiiStringari}, since we assume that bosons occupy a
single quantum state. The hydrodynamic equations for fermions can be
also put in a form of the Schr\"odinger-like equation by using the
inverse Madelung transformation
\citep{Dey98,Domps98,Grochowski17}. This is just a mathematical
transformation which introduces the single complex function instead of
density field and velocity field (which represents the potential flow)
used in a hydrodynamic description. Both treatments are equivalent
provided the velocity field is irrotational.

Introducing a condensed Bose field $\psi_B=\sqrt{n_B} \exp{(i \phi_B)}$ (with
$n_B=|\psi_B|^2$ and ${\bf{v}}_{\!B}=(\hbar/m_B) \nabla \phi_B$) 
and a pseudo-wavefunction for fermions $\psi_F=\sqrt{n_F} \exp{(i \phi_F)}$ 
(with $n_F=|\psi_F|^2$ and ${\bf{v}}_{\!F}=(\hbar/m_F)
\nabla \phi_F$) one gets 
\begin{eqnarray}
&& i \hbar \frac{\partial \psi_B}{\partial t} = H^{eff}_B \,\psi_B  \nonumber \\
&& i \hbar \frac{\partial \psi_F}{\partial t} = H^{eff}_F \,\psi_F    \,.
\label{eqmWD}
\end{eqnarray}  
The effective nonlinear single-particle Hamiltonians are given by
\begin{eqnarray}
H^{eff}_B &=& -\frac{\hbar^2}{2 m_B}\nabla^2 
 + g_B\, |\psi_B|^2 + \frac{5}{2} C_{LHY}\, |\psi_B|^3    \nonumber \\
&+&  g_{BF}\, |\psi_F|^2 + C_{BF}\, |\psi_F|^{8/3} A(w,\gamma)   \nonumber \\
&+&  C_{BF}\,|\psi_B|^2 |\psi_F|^{8/3}\, \frac{\partial A}{\partial \gamma} \frac{\partial \gamma}{\partial n_B}
  \,, \nonumber \\
H^{eff}_F &=& -\frac{\hbar^2}{2 m_F}\nabla^2 
+ \xi' \frac{\hbar^2}{2 m_F} \frac{\nabla^2 |\psi_F|}{|\psi_F|}
+ \frac{5}{3} \kappa_k |\psi_F|^{4/3}  \nonumber \\
&+&  g_{BF}\, |\psi_B|^2 + \frac{4}{3} C_{BF}\, |\psi_B|^2 |\psi_F|^{2/3} A(w,\gamma)   \nonumber \\
&+&  C_{BF}\,|\psi_B|^2 |\psi_F|^{8/3}\, \frac{\partial A}{\partial \gamma} \frac{\partial \gamma}{\partial n_F}  \,. 
\label{HamBF}
\end{eqnarray}
For the bosonic field a variety of quantized vortex 
states are possible, see \cite{Fetter01}, like for the single Gross-Pitaevskii 
equation. The bosonic wave function and the fermionic pseudo-wave function are 
normalized to the total number of particles in bosonic and fermionic components,
$N_{B,F} = \int d\mathbf{r}\, |\psi_{B,F}|^2$.  \\

\noindent
\textbf{Scaling up.}
Our simulations refer to the Bose-Fermi droplet consisting of bosonic
and fermionic atoms, as a model of cold white dwarf.
All quantities in the simulations are given in the code unit.
The  mass is expressed in $m_B$ -- the mass of bosonic atom 
(=133u in our case). 
A typical scattering length for atoms $a_B$ ($\simeq 5$ nm) represents the length unit 
and the time unit is given in $(m_B a_B^2)/\hbar$. 
The last one results from the Schr\"odinger equation ($E \propto \hbar^2\nabla^2/(2m)$ 
and $t \propto \hbar/E$).

Although the atomic droplet behaviour is simulated, the results can be scaled up 
to real astronomical sources.  To achieve this, let's adopt:
\begin{eqnarray}
r_{\rm astro} \ [\mathrm{m}] & = & a \, r_{\rm num} \ [\mathrm{m}] \\ 
m_{\rm astro} \ [\mathrm{kg}] & = & b \, m_{\rm num} \ [\mathrm{kg}] \\
t_{\rm astro} \ [\mathrm{s}] & = & \frac{ba^2}{c} \, \mathcal{T} t_{\rm num} \ [\mathrm{c.u.}] \ , \label{eq:time}
\end{eqnarray}
where $a, b$, and $c$ are unknown constants and c.u. means code unit. Coefficient 
$\mathcal{T} \equiv \mathrm{(m_B a_B^2)/\hbar} = 5.23 \times 10^{-8}$ s.
The droplet radius $r_{\rm BF} \simeq 1\mu$m $(\equiv r_{\rm num})$, and its mass
$m_{\rm BF} = 1042 \cdot 133$u $+ 100 \cdot 7$u $= 1047 m_B = 
2.31 \times 10^{-22}$ kg $(\equiv m_{\rm num})$.
By assuming that  $r_{\rm astro}$ and $m_{\rm astro}$ are 
the typical white dwarf radius and mass 
($0.01 R_{\odot},1 M_{\odot}$),
 we get $a = 6.98 \times 10^{12}$ and $b = 8.66 \times 10^{51}$.
Eq.~(\ref{eq:time}) actually leads to the relationship:
\begin{eqnarray}
\hbar & = & c \, \hbar_{\rm num} \ ,
\end{eqnarray}
where $\hbar = 1.05 \times 10^{-34}$ J\,s and $\hbar_{\rm num}$ is the
Planck constant used in the simulations.
Note, that in the case of the Bose-Fermi droplet studied 
experimentally, $a, b$, and $c$ constants have to equal 1.

Based on the Schr\"odinger equation, the energy must be scaled up according to:
\begin{eqnarray}
E_{\rm astro} \ [\mathrm{J}] = \frac{c^2}{b a^2} \, \mathcal{E} \, E_{\rm num} \ 
[\mathrm{c.u.}] \ ,
\end{eqnarray}
where $\mathcal{E} \equiv \mathrm{\hbar^2/(m_B a_B^2)} = 2.01 \times 10^{-27}$ J.
This relationship also applies to the potential energy and thus the gravitational
constant:
\begin{eqnarray}
G \ [\mathrm{m^3/(kg\,s^2)}] = \frac{c^2}{b^3 a} \, \mathcal{G} \, G_{\rm num} \ 
[\mathrm{c.u.}] \ ,
\label{eq:G}
\end{eqnarray}
where $G = 6.67 \times 10^{-11}$ and $\mathcal{G} \equiv 
\mathrm{\hbar^2/(m_B^3 a_B)}$ equals to $2.07 \times 10^{14}$, 
both values in units of m$^3$ kg$^{-1}$ s$^{-2}$.

To calculate $c$ constant, let's take that $G_{\rm num} M_{\rm BH \, num} =
1.93$ and $M_{\rm BH \, num}/m_{\rm BF} = 10$ (few periastron passage case). 
Note, that this ratio is also equal to $(\M/\m)_{\rm astro}$. Based on those values and
Eq.~(\ref{eq:G}) we get $c = 8.91 \times 10^{73}$. In the case of $\M = 10^4 \m$,
$c$ constant takes the value of $2.81 \times 10^{75}$.

Having $a$, $b$, and $c$ factors evaluated,
let's estimate the peak luminosity emitted during the third and the biggest 
mass loss (Fig.~\ref{accretion}). This passage results in the death of the WD-BH system.
Under the assumption that about 10\%\ of the white dwarf mass is stripped off by the black hole
and 10\%\ of the accreted mass is converted into radiation,
the total released energy is of the order of $10^{52}$ erg.
This energy is radiated in $\Delta t_{\mathrm{num}} \simeq 10^4$ [c.u].
It corresponds to  $\Delta t_{\mathrm{astro}} \simeq 2.5$ s while $\M = 10 \m$ and it reduces to $\sim 0.1$ 
s for $\M = 10^4 M_{\odot}$ (Eq.~\ref{eq:time}).

\bibliographystyle{naturemag}
\bibliography{BHWD_NATURE}

\noindent \\ \\
\textbf{Acknowledgements}
T.K., M.G., and M.B. acknowledge support from the (Polish)
National Science Center Grant number 2017/25/B/ST2/01943 and
M.N. acknowledges support from NSC Grant number
2016/22/M/ST9/00583. Part of the results were obtained using computers
at the Computer Center of University of Bialystok. \\

\noindent
\textbf{Author Contributions}  All authors made essential contributions to the work, 
discussed results, and contributed to the writing of the manuscript. The numerical
simulations were performed by T.K.. \\ \\

\end{document}